\documentclass[cpp,a4paper,fleqn%
]{w-art}
\usepackage{times,cite,w-thm}
\theoremstyle{plain}

\theoremstyle{definition}

\usepackage[]{graphicx}
\begin{document}
\DOIsuffix{theDOIsuffix}
\Volume{46}
\Month{01}
\Year{2007}
\pagespan{1}{}
\Receiveddate{XXXX}
\Reviseddate{XXXX}
\Accepteddate{XXXX}
\Dateposted{XXXX}



\title[Coulomb scattering in strong laser fields]{Fast electron generation by Coulomb scattering on spatially correlated ions in a strong laser field}


\author[S. Bauch]{Sebastian Bauch%
  \footnote{Sebastian Bauch \quad E-mail:~\textsf{bauch@theo-physik.uni-kiel.de},
                     }}
\address[\inst{1}]{Christian-Albrechts-Universit\"at zu Kiel, Institut f\"ur Theoretische Physik und Astrophysik, Leibnizstra\ss{}e 15, D-24098 Kiel}
\author[M. Bonitz]{Michael Bonitz}
\begin{abstract}
Electrons  colliding with spatially fixed ions in strong laser fields are investigated by solving the time-dependent Schr\"odinger equation.
Considering first simple one-dimensional model systems, the mechanisms and energy spectra of fast electrons are analyzed, starting from collisions on a single ion. By using these electrons as projectiles for a second and third collision, the maximum possible energy obtained can be significantly increased.  We then generalize the analysis to 2D systems where additional angular degrees of freedom lead to a drastic loss of efficiency. This problem can be overcome by introducing external confinements, which allow to focus the electrons and increase the intensity of high-energy electrons.
\end{abstract}
\maketitle                   





\section{Introduction}

Electron acceleration, or more general acceleration of charged particles, is of importance for many applications in science and technology. The traditional way uses strong electro-magnetic fields from large, super-conducting coils. More recently, new technologies relying on (non-linear) optical effects in strong laser fields became available, leading to a dramatic decrease in the needed sizes of experiments to finally give them a ``table-top'' character. On the one hand there is enormous progress in higher harmonics generation in strong laser fields, where the ionization of rare gas atoms and the subsequent rescattering mechanisms lead, besides the well-known production of (ultra-short) intense UV/XUV radiation, also to generation of fast electrons. On the other hand, wake field acceleration, where the electrons ``surf'' on the laser wave, offers a new possibility to create high-energy electrons on small spatial scales \cite{katsouleas}.
In plasmas, the dominating mechanism of energy transfer from a laser field to an electron is inverse bremsstrahlung occurring in electron-ion collisions. Here, the electron absorbs photons from the electro-magnetic field during a collision with an ion. This subject has been investigated within a quantum kinetic theory in \cite{kremp, bonitz}, and numerical results showing bunches of fast electrons were found in \cite{haberland}.
The above mentioned systems are macroscopic with random particle arrangements, e.g. in plasmas or gas-phase situations which strongly reduces the acceleration efficiency. An alternative concept is the microscopic electron scattering on spatially correlated ions. Here, statistical effects, as e.g. avaraging of inter-ionic distances, are eliminated and the underlying fundamental processes become visible.

By the group of Kull et al. electrons colliding with single ions in the presence of strong laser fields were investigated in detail\cite{kull} by means of solving the time-dependent Schr\"odinger  equation (TDSE). The occurring distributions of fast electrons being accelerated by the laser field were explained within classical models which often describe strong field physics surprisingly well ---the most prominent example is the simple-mans theory for higher-harmonics generation\cite{corkum, corkum89, gallagher}. By using the accelerated electrons as projectiles for a second collision with an additional absorption of energy, a resonance behavior regarding the inter-ionic distance with an enormous increase in kinetic energies of scattered electrons could be found in one-dimensional model systems\cite{goerlinger}. In a later work, multi-dimensional solutions of the TDSE allowed for the accurate estimation of angular distributions of scattered electrons on single ions\cite{rascol}.

Despite the exhaustive investigation of these processes, some questions remain open:
can the idea of resonant energy absorption by scattering on spatially correlated ions pushed further, having in mind the possible experimental use as a source of fast electrons. And secondly, is it, in general, possible to extend the one-dimensional results in \cite{goerlinger} for correlated scattering on ion pairs to the multi-dimensional case, where wave-packet spreading and additional angular degrees of freedom  lead to a substantial decrease in intensity of fast electrons.
 It is the intention of this work to answer these questions and to propose methods to enhance the intensity of fast, optimally scattered electrons.

The paper is organized as followed: in the first part we introduce the used tools and methods, mainly based on TDSE simulations of wave-packet scattering. Then we discuss the mechanism of energy absorption by studying one-dimensional electron scattering on single ions as well as on two spatially correlated ions and generalize the setup to three ions allowing for more (resonant) scattering processes.
In the next part we analyze two-dimensional Coulomb collisions on one and two ions and investigate the occurring angular distributions of scattered electrons in detail. Finally, we propose a method to increase the intensity of fast electrons by introducing additional focussing confinement potentials.

\section{Method}
 
\begin{figure}
 \sidecaption
 \includegraphics[width=0.35\textwidth]{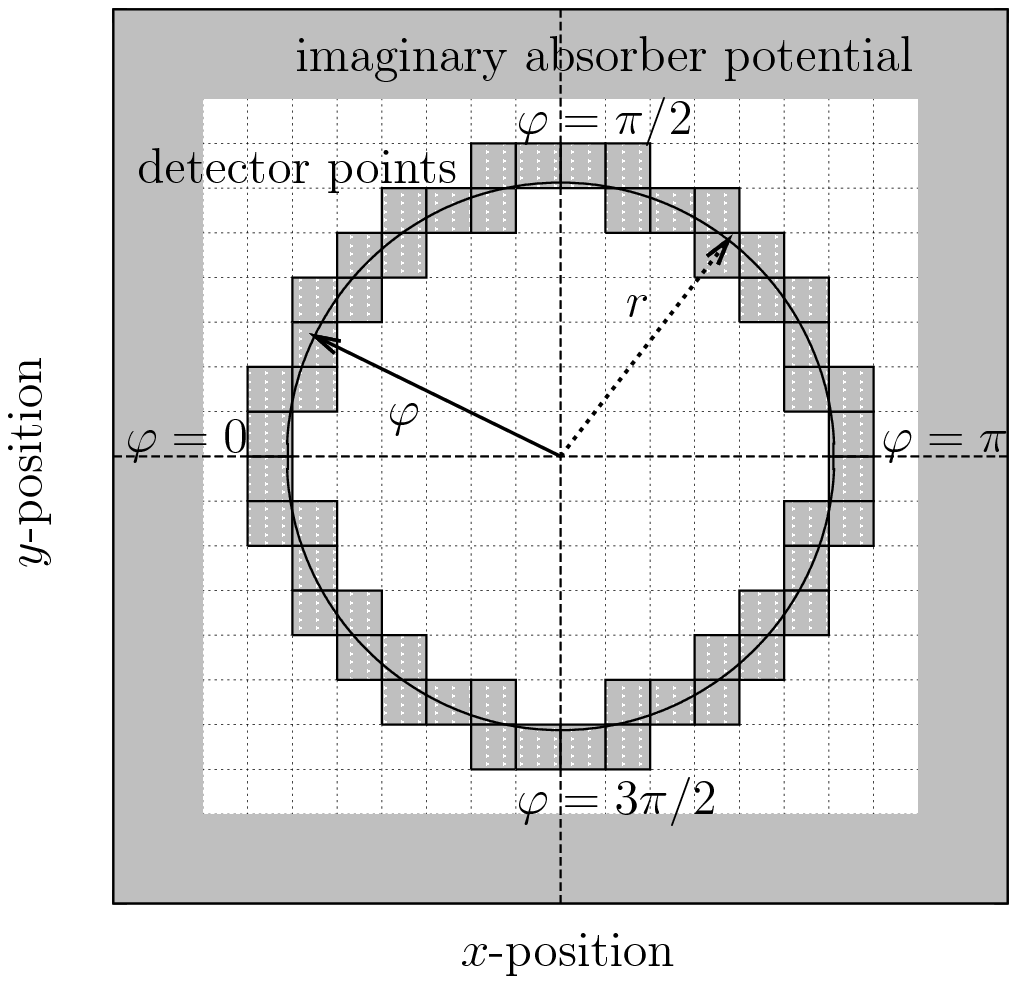}
 \caption{Numerically implemented spherical detector  with radius $r$ for (fast) electrons, the ion is at the origin. A large number of grid points (typically $>10^4$) is used to calculate the angle-resolved electron spectra, see \cite{bauch} for details. An imaginary absorbing potential (\emph{optical} potential) behind the detector removes detected electrons from the system to avoid reflections at the edge of the simulation box. }
 \label{fig:detector}
\end{figure}

The scattering of electrons on ions in the presence of a strong laser field is quantum mechanically accurately described by the (one-particle) time-dependent Schr\"odinger equation in spatial coordinate representation (throughout this work atomic units [a.u.] are used, $\hbar=m_e=e=1$),
\begin {equation}
 i \frac{\partial}{\partial_t} \Psi (\boldsymbol{r},t)=  \left (- \frac{1}{2} \Delta + V(\boldsymbol{r},t) \right ) \Psi(\boldsymbol{r},t)\; ,
 \label{eq:tdse}
\end {equation}
where the ion is assumed to be fixed in space $(m_{\textup{ion}}\rightarrow \infty)$.
The potential term $V(\boldsymbol{r},t)$ splits up into a time-independent ionic part, $V_{\textup{ion}}(\boldsymbol{r})$, and a time-dependent part, $V_{\textup{laser}}(\boldsymbol{r},t)=E_0 x \cos(\omega t)$, describing the laser field in dipole approximation. $E_0$ is the field amplitude and $\omega$ the photon energy. The ionic potential is given by  a regularized Coulomb potential,
\begin {equation}
 V^{\textup{1D}}(x) = -\frac{Z}{\sqrt{x^2+\kappa^2}} \hspace{1cm} \textup{resp.} \hspace{1cm}  V^{\textup{2D}}(x,y) =-\frac{Z}{\sqrt{x^2+y^2+\kappa^2}} .
 \label{eq:coulomb}
\end {equation}
The charge number $Z$ is chosen to be one, hence giving, e.g., a model for elementary processes in a totally ionized hydrogen plasma.  
The regularization by $\kappa$ of the potential is needed to prevent the singularity of the Coulomb potential. This parameter strongly influences the scattering properties of the potential, as, e.g., the quantum mechanical reflection and transmission coefficients. To compare with reference \cite{kull}, we used $\kappa=0.1$ throughout this work. 

Following \cite{kull} we transform the TDSE, equation~\eqref{eq:tdse}, into the Kramers-Henneberger (KH) frame of reference \cite{kramers,henneberger}, in which the laser  field leads to a time-dependent shifted ionic potential term, $V^{\textup{KH}}[\boldsymbol{r}^{\textup{KH}}(t)]$ (the observer ``sits'' on the electron wiggling in the laser field, details are to be found in \cite{kull,kramers,henneberger}).

We solve equation~\eqref{eq:tdse} in the KH-frame numerically on large spatial grids in one and two dimensions utilizing a Crank-Nicolson procedure in combination with the split operator technique which gives access to the multi-dimensional solution. Details of the method and its implementation can be found in \cite{numrec, hobokenbook}. The angle-resolved energy spectra of the (scattered) electronic wave function are obtained using an experiment-like detector described in \cite{bauch}, briefly sketched in figure~\ref{fig:detector}. Here, the energy spectrum is computed utilizing a fast Fourier transform of the wave function on each detector point with respect to the time $t$, which gives the energy spectrum of potential free particles. The advantage of this procedure, in comparison to \cite{kull}, where the whole wave function $\psi(\boldsymbol{r})$ is transformed into momentum space, is the possibility to introduce an optical potential, which removes scattered electrons from the grid after their detection, e.g. \cite{neuhauser, vibok}. This allows for significantly smaller grids and reduces the numerical effort in more than one spatial dimensions enormously.

As initial condition, we use electronic wave packets of Gaussian form (here the 1D case is shown, the generalization to 2D is straightforward),
\begin {equation}
  \Psi(x,t=0) = \frac{1}{\sqrt{2 \pi \sigma}} \exp \left (  - \frac{(x-x_0)^2}{2\sigma^2}\right ) \exp(i k_0 x) \; .
 \label{eq:wavepacket}
\end {equation}
The width $\sigma$ determines, via Heisenberg's principle, the energy resolution. $x_0$  and $k_0$ are the initial position and momentum.
The wave packet is placed in front of an ion at a position $x_0=-k_0 \cdot N_{\textup{laser}} \pi/\omega$ and travels with an initial momentum $k_0$ in direction of the scattering setup. $N_{\textup{laser}}$ denotes the number of laser cycles during the collision process, typically $N_{\textup{laser}}=16$ is used, which was found to be not a sensitive parameter. The width of the wave packet, $\sigma$, is chosen to be large enough, to allow for scattering at any phase of the laser field and to allow for sufficiently high energy resolution. In most calculations $\sigma=0.4\cdot x_0$ was used.

\section {1D-Coulomb-Scattering Results}
In the following, we will discuss the scattering of electronic wave packets, cf. equation~\eqref{eq:wavepacket}, on spatially fixed ions in one dimension.  Figure~\ref{fig:1d_single}, center graph, shows the initial wave packet and the corresponding final state for a single-ion scattering setup. Additionally, both detector positions (left and right of the Coulomb potential in the central region) are indicated by crosses, which is the natural one-dimensional simplification of the setup displayed in figure~\ref{fig:detector}. 

\subsection{Electron scattering on one ion}
\label{sec:1d_1ion}

\begin{figure}
  \begin {minipage}{0.73\textwidth}
   \includegraphics[width=0.98\textwidth]{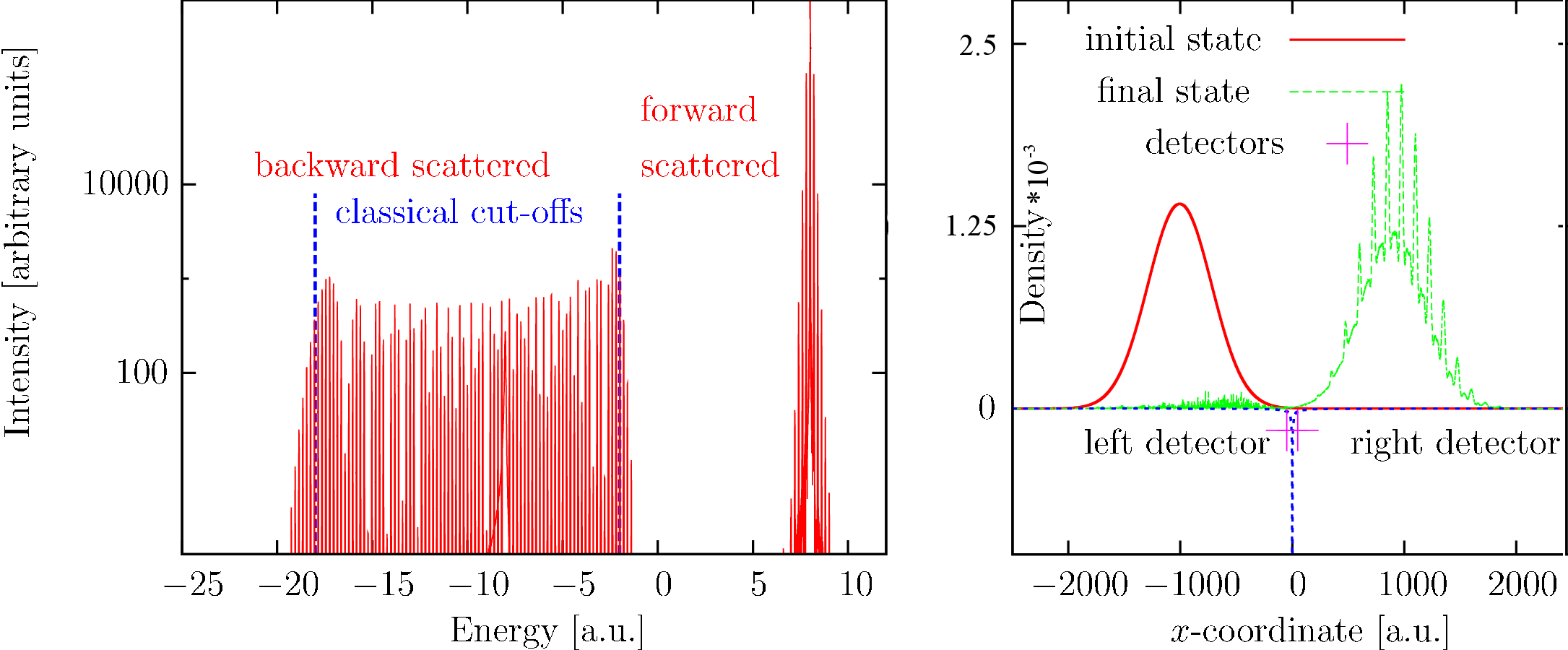}
\end {minipage}
\begin{minipage}{0.26\textwidth}
  \includegraphics[width=0.87\textwidth]{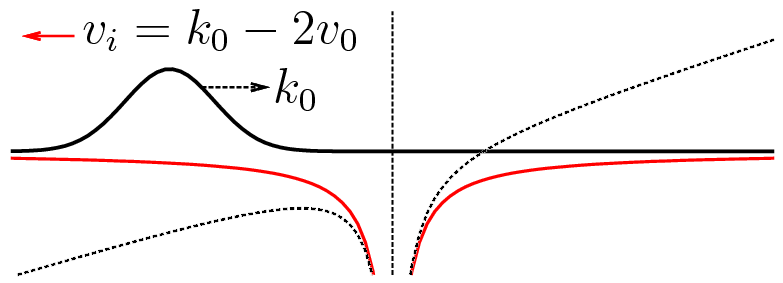} (a)\\
  \includegraphics[width=0.87\textwidth]{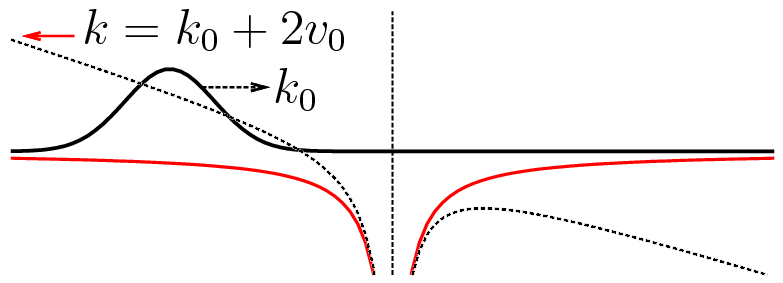} (b)\\
  \includegraphics[width=0.87\textwidth]{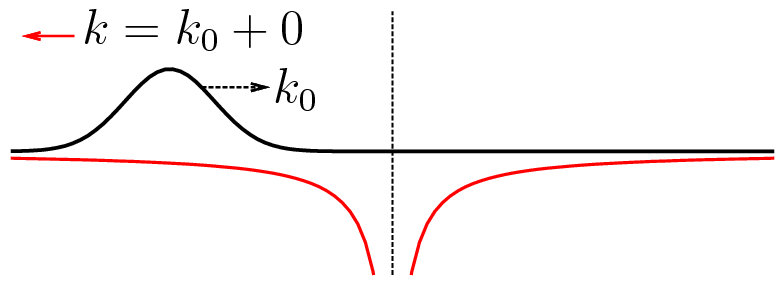} (c)\\
 \end {minipage} 
\caption{(color online) 1D Coulomb scattering on a single ion in a strong laser field, $\omega=0.2$ and $E_0=0.2$. The scattering setup, showing both detectors for electrons (+) and the initial and final electronic densities, are given in the central plot. The left figure shows the energy spectrum of the scattered electrons for an initial momentum of $k_0=4.0$. Positive (negative) energies indicate scattering to the right (left) of the ion. Figures (a-c) on the r.h.s sketch the (classical) energy absorption mechanism where the grey line indicate the direction of the laser potential \emph{before} time of scattering $t_s$ (see text for details). This leads to the classical cut-off energies marked by (blue) dashed lines in the left-most graph.}
\label{fig:1d_single}
\end{figure}
To check the concept of our detector setup and to compare with previous investigations, we first look at scattering on a single ion. The corresponding results for $k_0=4.0$ in a strong, linearly polarized laser field of $\omega=0.2$ and $E_0=0.2$ are shown in figure \ref{fig:1d_single}, left graph, and agree very well with \cite{kull}. The absorption of single photons from the (classical) dipole laser field manifests in the occurrence of a sequence of equidistant peaks in the energy spectrum, accurately separated by the photon energy $\omega$. Clearly, a distribution of fast electrons being reflected at the ion (backward direction is indicated with negative energies in the spectra), is observed in the electron spectrum exhibiting a large plateau with two significant cut-off energies. In forward direction (positive energies), the peak of the transmitted wave packet at an energy of $k_0^2/2=8.0$ is clearly visible.
According to the classical investigation in \cite{kull}, the cut-off energies of the plateau in the energy domain can be explained by using the momentum conservation law, $k+P_{\textup{E}}=k_0+P_{\textup{E}}$, involving the momentum before the collision, $k_0$, and after the collision, $k$. Here, $P_{\textup{E}}=v_0 \cos \omega t_s$ denotes the quiver momentum at time of scattering, $t_s$, and $v_0=E_0/\omega_0$ is the amplitude of the quiver velocity of the electron in the laser field. Transformation of this formula gives for the minimum and maximum transfered momentum to the electron ($\cos \omega t_s^* = \pm 1$):
$k_{\textup{max}}=-k_0-2v_0$ and $k_{\textup{min}}=-k_0+2v_0$, which corresponds to kinetic energies of $E_{\textup{min,max}}=k_{\textup{min,max}}^2/2=(k_0\pm 2v_0)^2/2$.
Figures \ref{fig:1d_single} (a-c) schematically show the origin of the significant cut-offs in energy. Plotted is the ionic potential bent by the (strong) laser potential for three typical situations during the cycle \emph{before} the scattering. Recall that the potential is phase shifted with respect to the ponderomotive motion of the electron.  In the first situation, (a), the electron is decelerated before the collision, whereas in the second case, (b), the electron is accelerated towards the ion. Of course, any phase of the field inbetween both extrema is possible, which explains the  plateau distribution.

\subsection{Scattering on an ion pair}
\label{sec:1d_2ions}
\begin{figure}
 \begin {minipage}{0.45\textwidth}
 \includegraphics[width=0.98\textwidth]{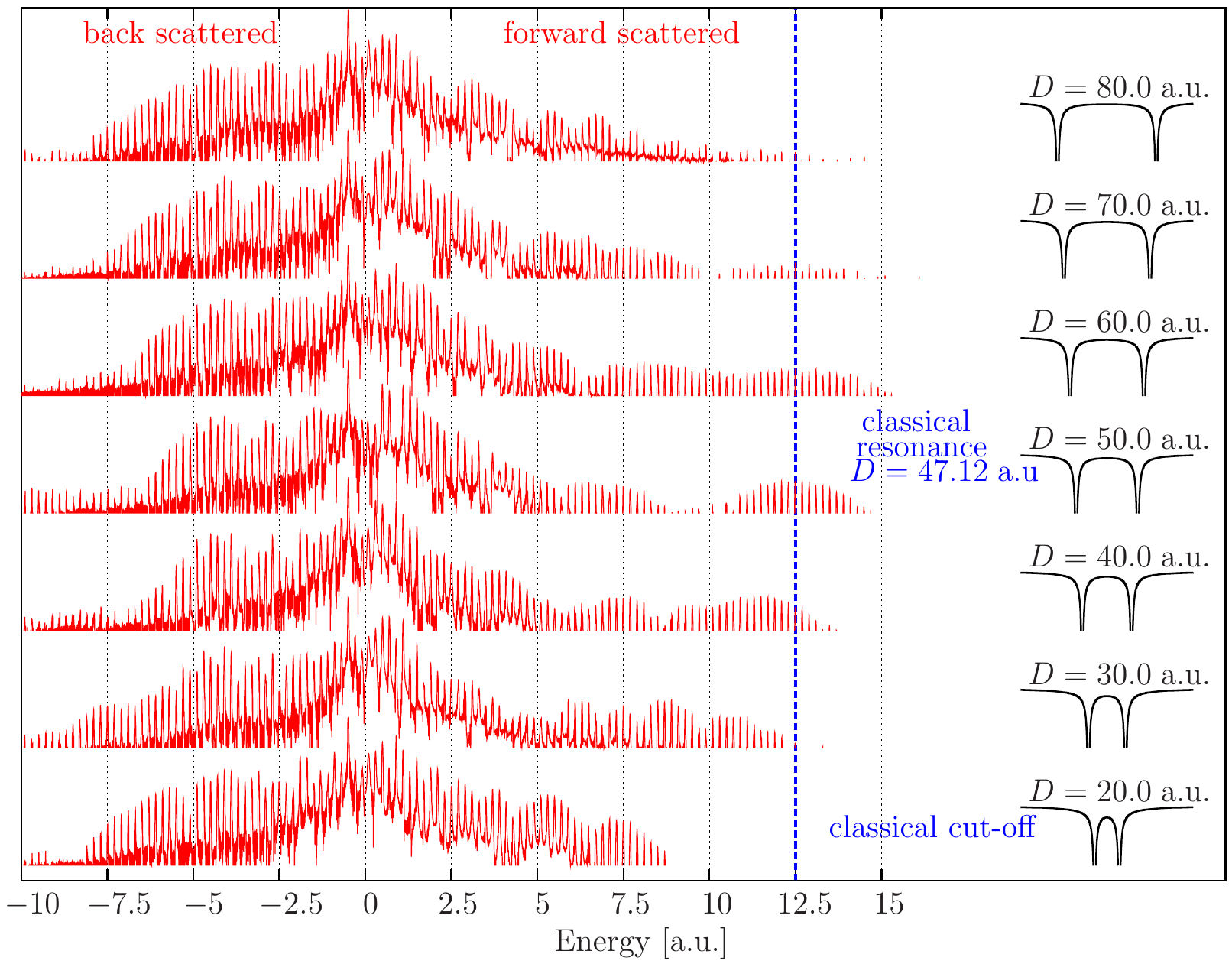} 
 \end {minipage}
 \begin{minipage}{0.26\textwidth}
  \includegraphics[width=0.87\textwidth]{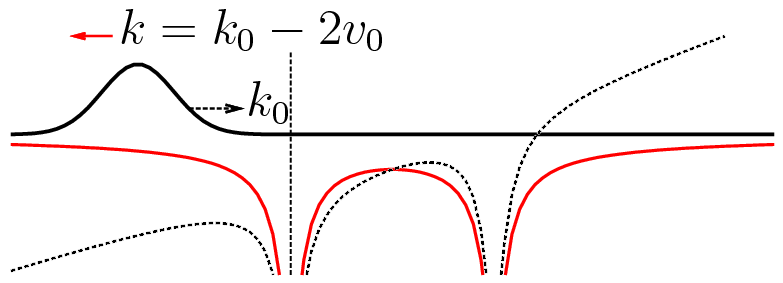} (a)\\
 \includegraphics[width=0.87\textwidth]{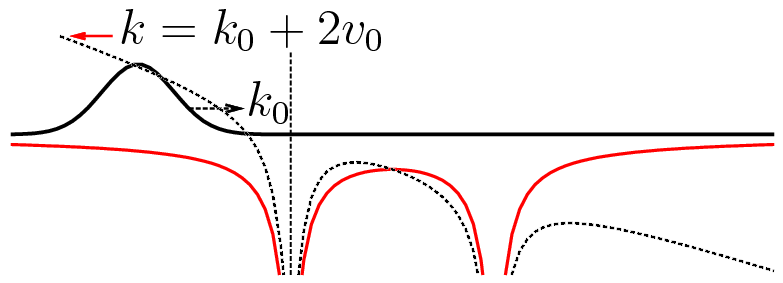} (b)\\
 \includegraphics[width=0.87\textwidth]{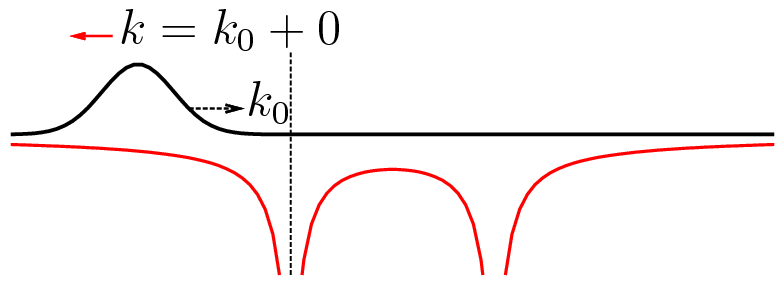} (c)\\
 \end{minipage} 
\begin{minipage}{0.26\textwidth}
\includegraphics[width=0.87\textwidth]{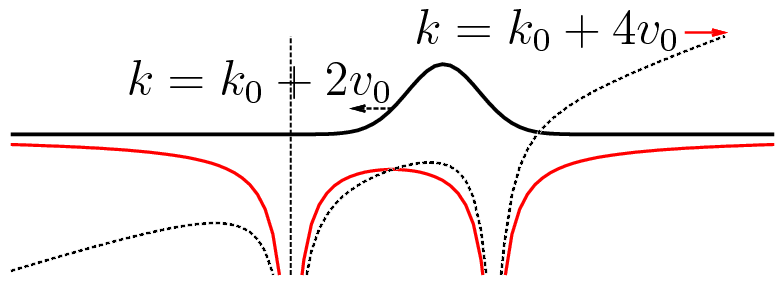} (d)\\
\includegraphics[width=0.87\textwidth]{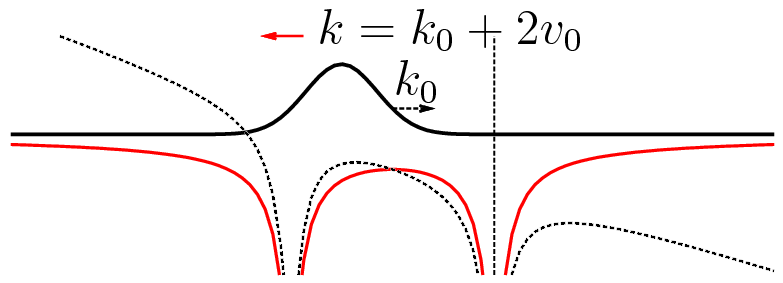} (e)\\
\includegraphics[width=0.87\textwidth]{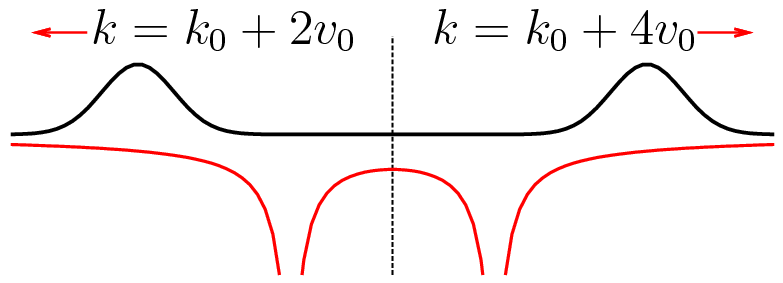} (f)\\
\end{minipage}
 \caption{(color online) 1D Coulomb scattering on two ions separated by a distance $D$ in a strong laser field, $\omega=0.2$ and $E_0=0.2$. The left figure shows the occurring electron energy distributions after a collision with $k_0=1.0$ by variation of $D$. The corresponding energy absorption mechanisms are sketched in figures (a-f). The formation of the left plateau cut-off (negative energies in the spectra) is displayed by figures (a-c), the cut-off for scattering to the right (positive energies) by resonant scattering between both ions by figures (d-f). The grey lines indicate the direction of the laser potential \emph{before} time of scattering, see text. }
\label{fig:1d_2ions}
\end{figure}

We now generalize the results to an arrangement of two ions, which are spatially separated by a fixed distance $D$ as sketched in figure~\ref{fig:1d_2ions}. According to \cite{goerlinger} a resonant increase in energy of scattered electrons can be found if the distance between both ions is chosen according to $D=(k_0+2v_0) \cdot \pi/\omega$, with a maximum cut-off energy of $E_{\textup{max}}=(4 v_0 + k_0)^2/2$. The results of TDSE simulations with variation of the distance $D$ are given in figure~\ref{fig:1d_2ions}. The underlying mechanisms leading to a distribution of fast electrons in both directions (backward and forward scattered) is displayed in figures~\ref{fig:1d_2ions}(a-f). As in the previous case, the direction of the ponderomotive momentum, indicated by the laser potential before the time of collision in the graphs (a-f), decides about the particle's energy gain during the collision. Upon scattering on the first ion, the backward scattered fraction possesses the energy distribution of the singly scattered electrons (a-c). But, by scattering on the second ion, the backward scattered electrons can again acquire energy during a second collision (d, e). If the field now changes its direction during both collision, both scattering processes lead to a maximum absorption of quiver momentum from the laser field, which is transfered into lateral motion of the electron. Hence, the forward scattered fraction of electrons exhibits an extended double-plateau structure in the energy spectra.

\subsection{Scattering on three ions}
\begin{figure}
  \begin{minipage}{0.59\textwidth}
 \includegraphics[width=0.98\textwidth]{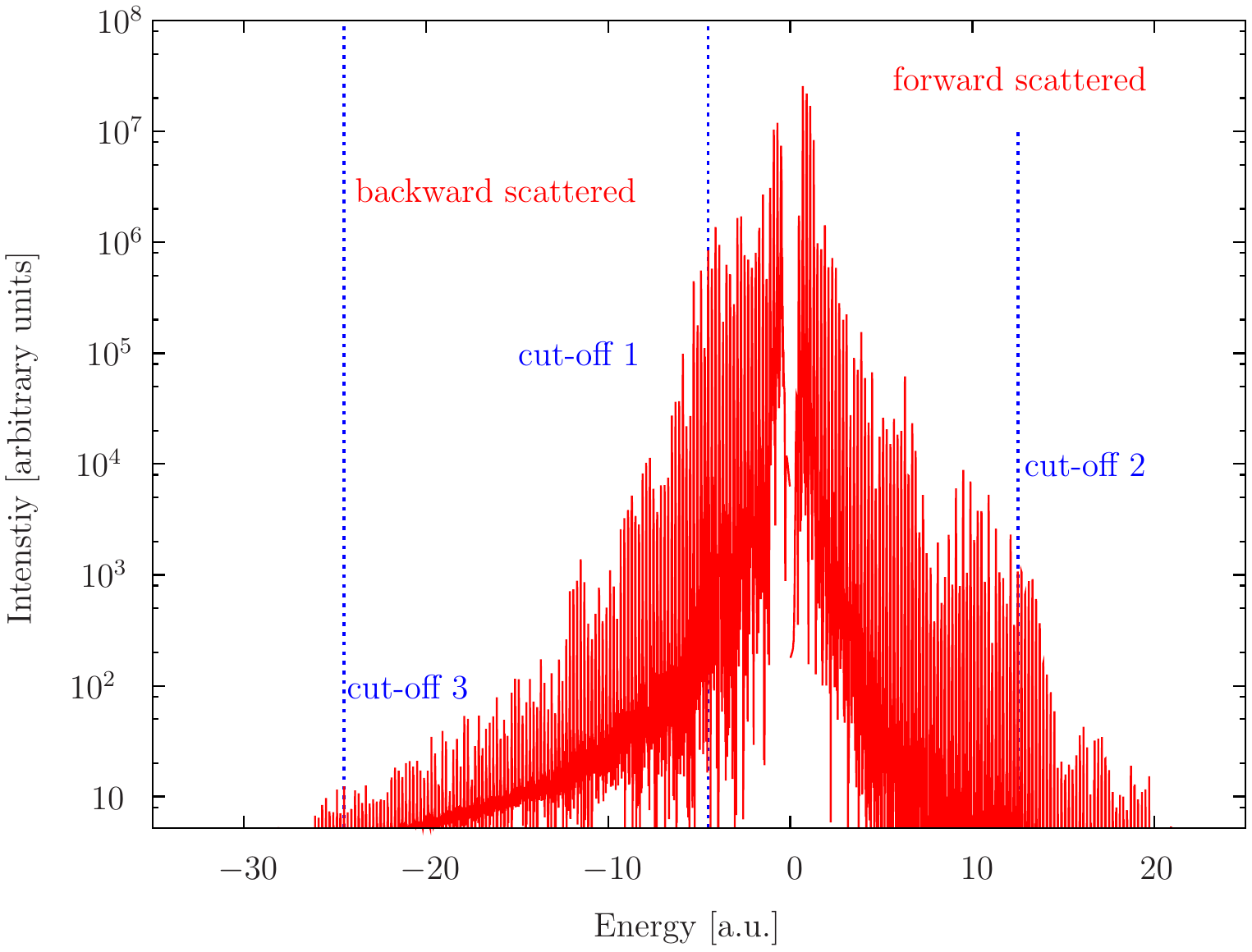}
  \end {minipage}
  \begin{minipage}{0.39\textwidth}
    \includegraphics[width=0.98\textwidth]{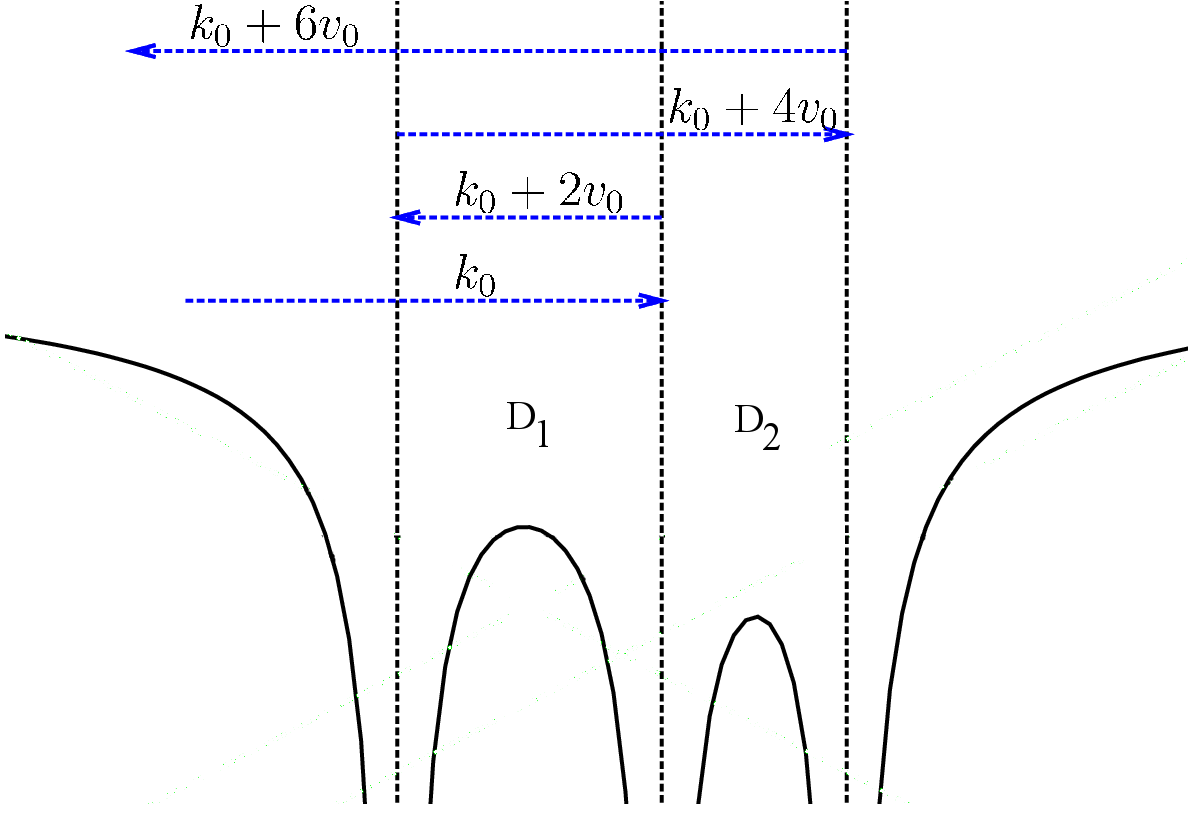}
  \end{minipage}
  \caption{(color online) 1D Coulomb scattering on three ions at distances $D_1=47.12$ and $D_2=31.4$ allowing for resonant absorption of energy from the laser field, see text for details. The right picture shows the corresponding processes leading to three significant cut-off energies in the spectrum, which is given in the left graph. Again, positive (negative) energies indicate scattering to the right (left). All other parameters are the same as in figure~\ref{fig:1d_2ions}.}
\label{fig:1d_3ions}
\end{figure}

To further increase the energy of scattered electrons, we now use the fast electrons of the previous described setup of two ions to let them collide with a third ion, which is placed at a distance $D_2$ allowing for another resonant absorption of energy. Analogously to the previous case, a resonance condition involving both distances can be derived, cf. right part of figure~\ref{fig:1d_3ions}.
The first scattering process takes place at the \emph{second} ion. Then, according to the double-scattering setup, the next ion has to be placed at a distance $D_1=(k_0+2v_0) \pi/\omega$ in \emph{front} of the first collision partner where a second scattering event happens. The faster electrons now travel towards a third ion, which is at a distance of $D_2=2v_0 \pi /\omega$, and are reflected with increased energy a third time. These electrons now have to pass the whole setup again and exit the system to the left (backward direction, negative energies in figure~\ref{fig:1d_3ions}) with an energy of up to $E_{\textup{max}}=(k_0+6v_0)/2$. 
The resulting electron energy spectrum of a TDSE simulation for such a process and the expected (classical) cut-off energies are shown in figure~\ref{fig:1d_3ions}. The signal of the three-fold scattered electrons forming the third plateau is very weak, since many processes occur simultaneously and parts of the wave function undergo several collisions at all types of combinations of laser field phases and only a very small fraction is accelerated optimally. Hence, the simulations have to be performed with high resolution on especially fine numerical grids as well in space as in time. For larger setups consisting of four or even more ions in a chain, the resonance conditions can of course be derived, but it is nearly impossible to detect the corresponding electrons forming the expected plateaus of resonantly scattered electrons in the energy domain.

\section {2D-Coulomb-Scattering Results}
As demonstrated above, the one-dimensional model allows for a computationally fast access to scattering processes in (strong) laser fields and the corresponding distributions of fast electrons and a simple understanding of the physical processes. Nevertheless, it is not able to make robust predictions for experiments because the intensity of high-energy electrons is clearly overestimated since it neglects the additional angular degrees of freedom and the spatial spreading of the electronic wave packets in a real system. We therefore switch now to 2D model systems which have been demonstrated to give access to accurate energy spectra of multi-photon processes and the corresponding angular distributions of photoelectrons\cite{bauch} and are expected to incorporate the dominant phenomena occurring in real systems.

\subsection{Angle-resolved electron scattering on one ion}

\begin{vchfigure}
\sidecaption
   \includegraphics[width=0.49\textwidth]{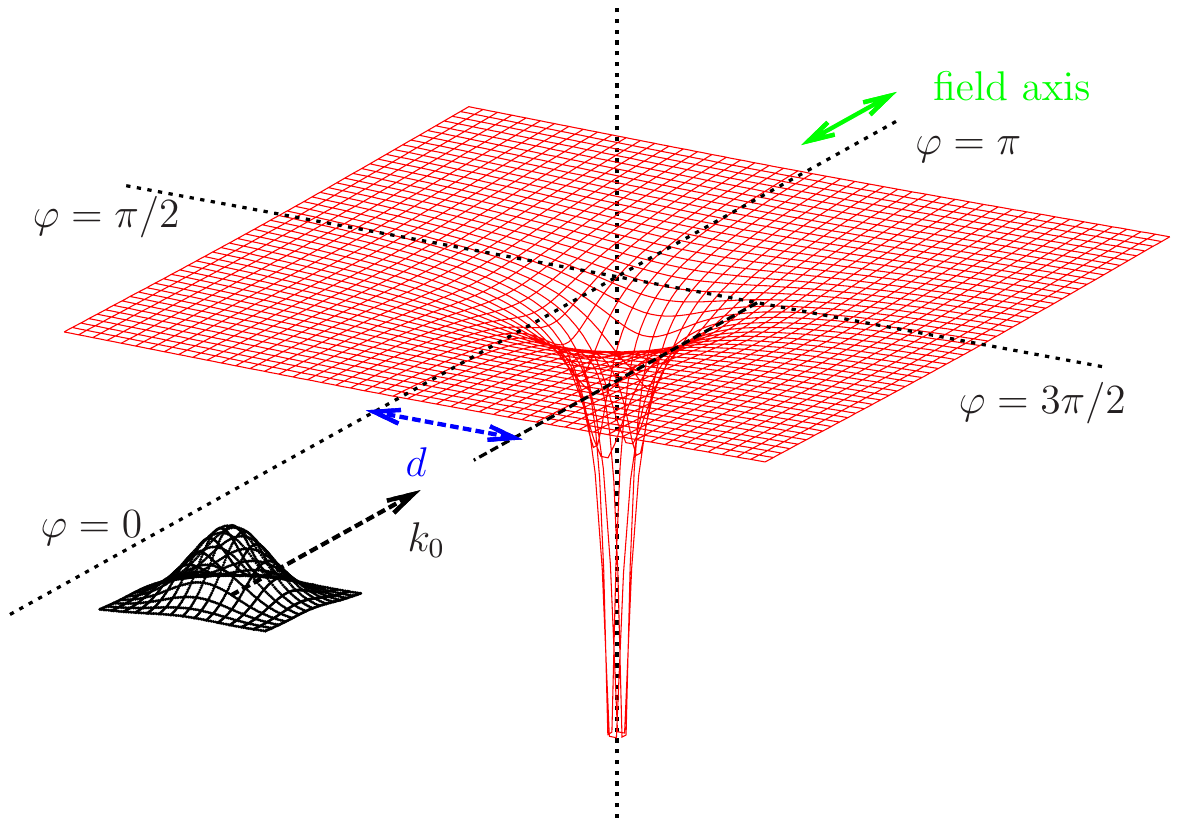}
    \caption{(color online) 2D Coulomb scattering on one single ion. The electronic wave-packet collides centrally ($d=0$) with the ion (center region). The initial momentum $k_0$ is parallel to the field polarization axis of the laser.}
\label{fig:2D_geometry}
\end{vchfigure}

\begin{figure}
 \begin{minipage}{0.49\textwidth}
  \includegraphics[width=0.98\textwidth]{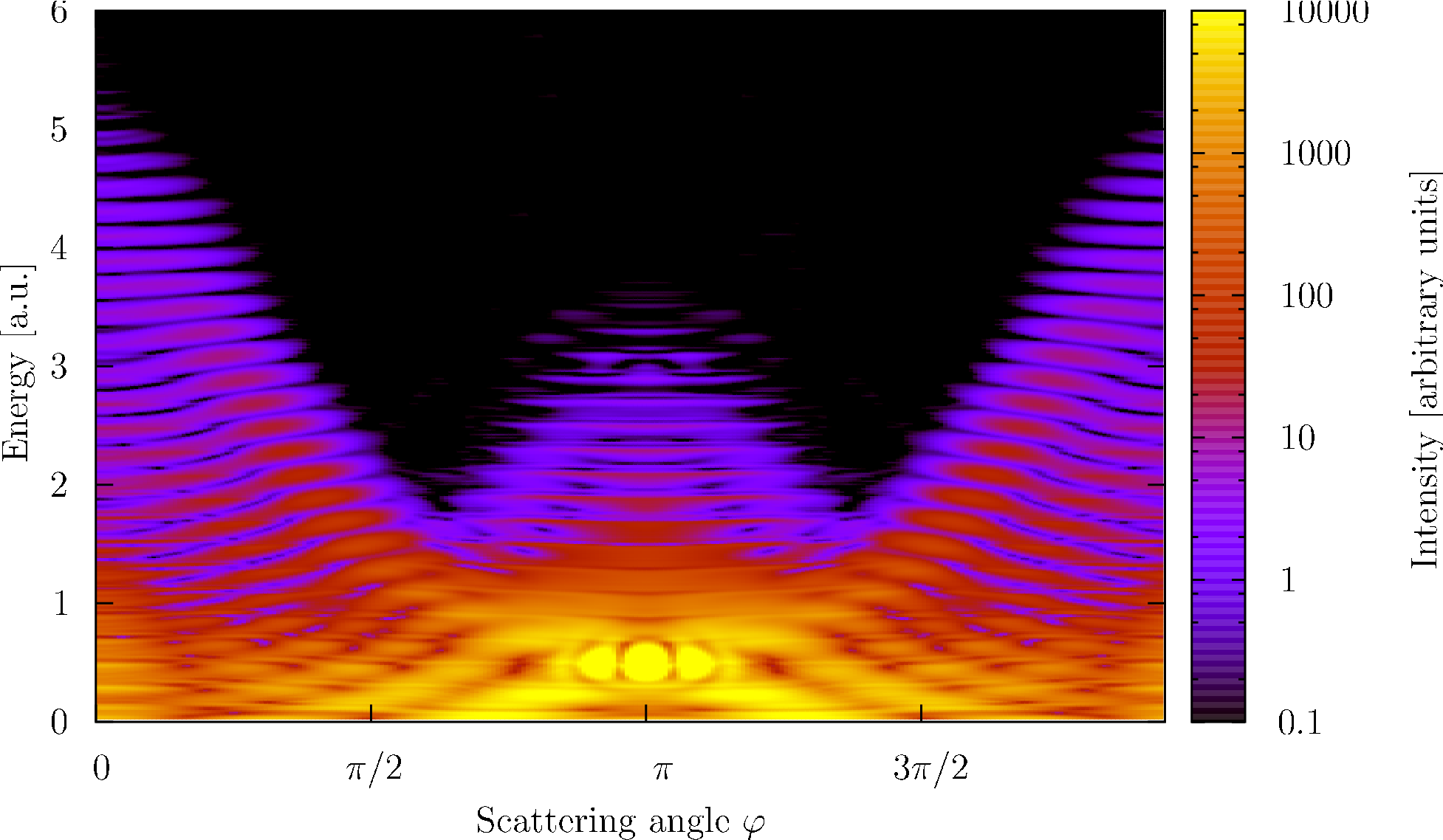}  
 \end{minipage}
  \begin{minipage}{0.49\textwidth}
   \includegraphics[width=0.98\textwidth]{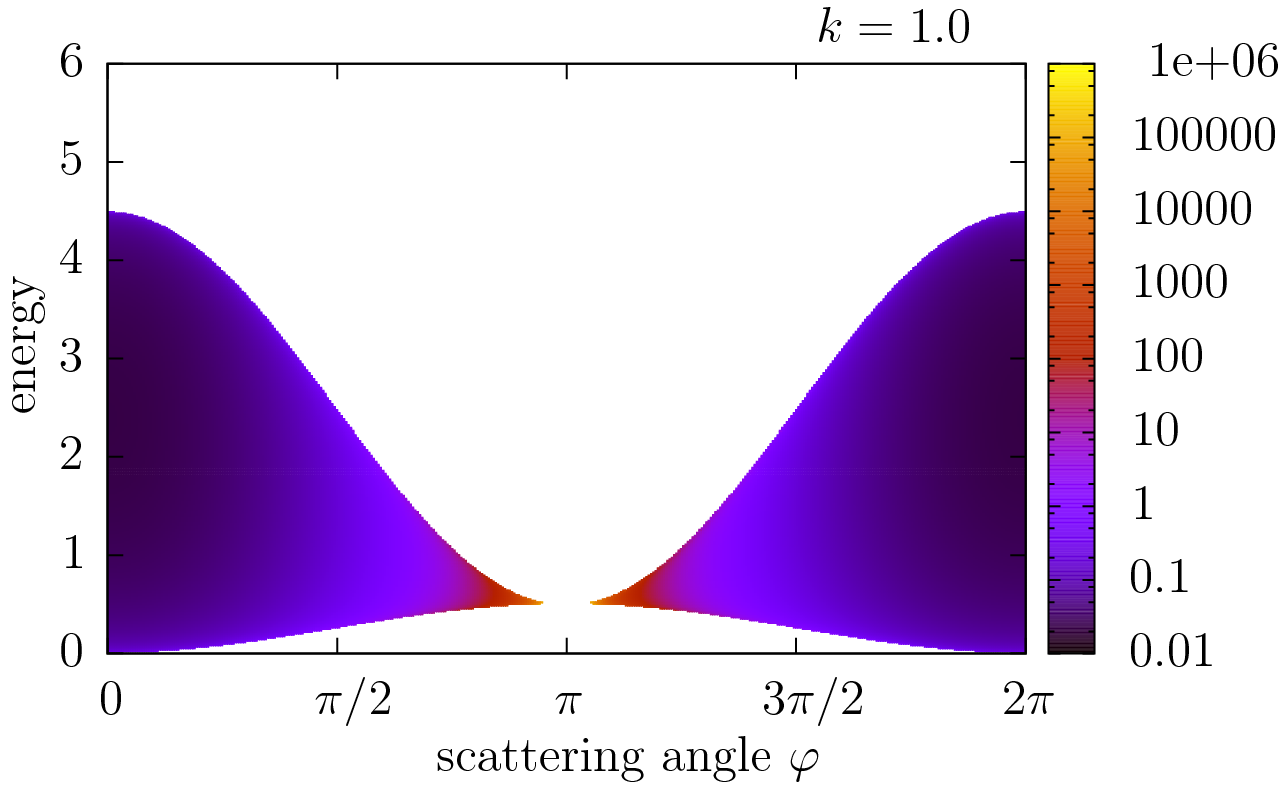}
  \end{minipage}
\caption{(color) 2D Coulomb scattering on one single ion in a strong laser field, $\omega=0.2$ and $E_0=0.2$ with an initial momentum of $k_0=1.0$. The left graph shows the energy distribution by solving the TDSE using the detector method. The laser field is polarized along the $x$-axis ($\varphi=0, \pi$). The right graph shows the  classically estimated spectrum of backward scattered (fast) electrons according to the classical model of instantaneous Coulomb collisions \cite{kull}. The scattering angle of $\varphi=0, 2\pi$ corresponds to backward scattering whereas $\varphi=\pi$ denotes forward  (in direction of $k_0$) scattering.}
\label{fig:2d_1ion}
\end{figure}

First, we consider electron scattering on one ion. As sketched in figure~\ref{fig:2D_geometry}, a wave packet with initial momentum $\boldsymbol{k}_0$ travels towards an ion and scatters centrally ($d=0$). The Gaussian distribution of the initial wave packet assures, that this parameter has no influence on the final result besides a possible intensity shift in the angular distribution of scattered electrons in accordance with $d$ \cite{bauchdiplom}. The direction of $\boldsymbol{k}_0$ is parallel to the field axis of the strong laser field ($\omega=0.2, E_0=0.2$, linearly polarized along the $x$-direction). Then, the scattered part of the wave function is detected, cf. figure~\ref{fig:detector}, and removed from the system by the optical potential. The corresponding angle-resolved energy spectrum of such a process ($k_0=1.0$) obtained by solving the TDSE is shown in figure~\ref{fig:2d_1ion}. One observes a richly structured angular dependence of the scattered electronic wave packet. The prominent peak-like energy distribution (well separated by the photon energy $\omega$) indicates the absorption of single photons from the laser field. We point out the similarity to angle-resolved above threshold ionization (ATI) spectra\cite{bauch}, besides an asymmetry induced by the initial momentum $\boldsymbol{k}_0$. This analogy between ATI and fast electron generation was already observed in 1D systems \cite{kull}.
As in the one-dimensional case, the large peak at $\varphi=\pi$ stems from the wave packet travelling in forward direction. The angle-dependent pattern originates from an intereference behavior at the Coulomb potential (in analogy to the well-known single slit experiment of quantum mechanics). Both classical cut-off energies in the backward ($\varphi=0,2\pi$) and forward ($\varphi=\pi$) direction are found in correspondence to the one-dimensional model.

Additionally, the angle-resolved energy spectrum can be calculated within the classical instantaneous Coulomb collision model presented in detail in\cite{kull}. The right part of figure~\ref{fig:2d_1ion} shows the result of such an investigation (cf. equations~(48) and (49) in \cite{kull}). Only the distribution of electrons which contributes to the high-energy part in backward direction is displayed which is to be compared to the distribution obtained within our TDSE simulations. Again, the classical explanation reproduces the accurate quantum calculation very well, except for pure quantum-effects (e.g. the absorption of photons from the laser field). For the foward scattered part, in direction of $\varphi=\pi$ a similar calculation can be performed, such that the second cut-off region in the TDSE spectrum and its shape are also explained within the classical model.

\subsection{Electron scattering on two ions}
\begin {figure}
 \begin{minipage}{0.4\textwidth}
  \includegraphics[width=0.98\textwidth]{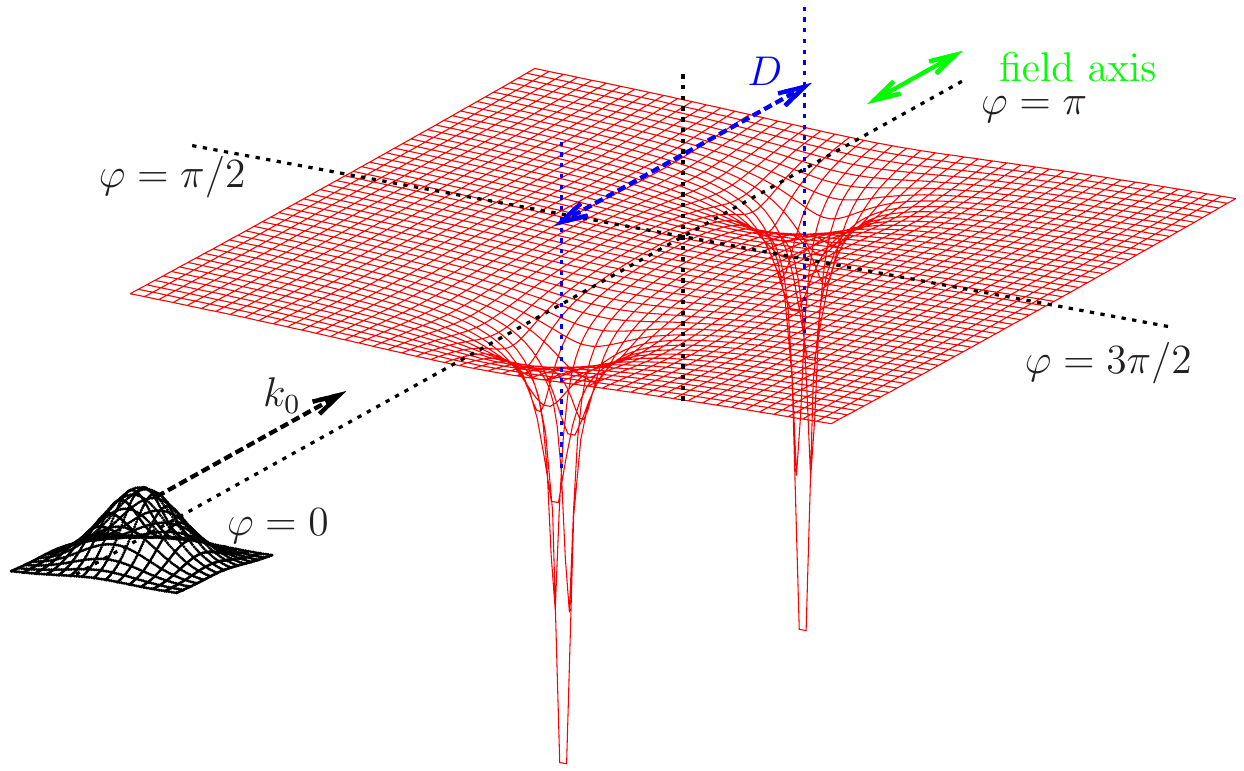}
 \end{minipage}
\begin{minipage}{0.59\textwidth}
   \includegraphics[width=0.98\textwidth]{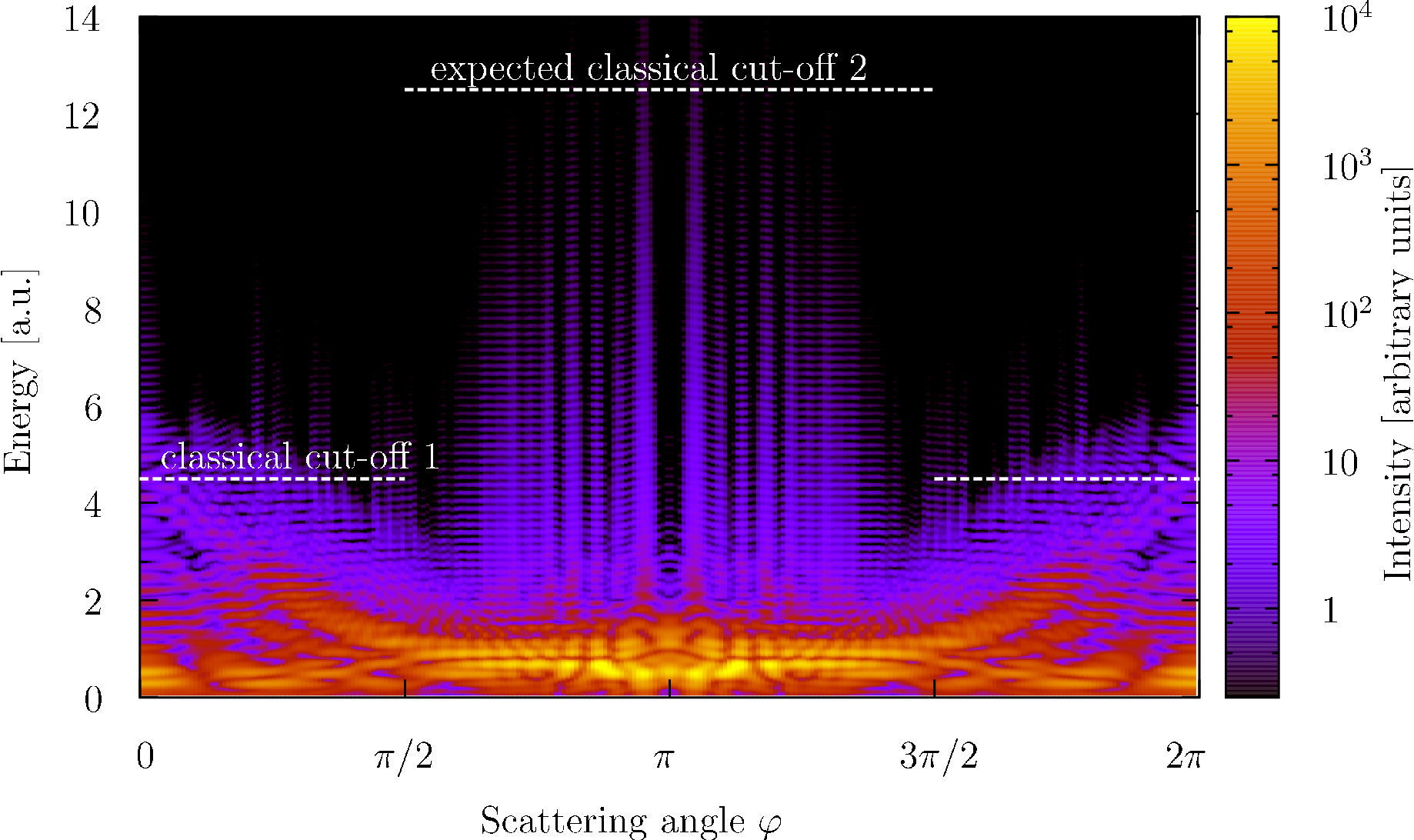}  
 \end{minipage}
 \caption{(color) 2D Coulomb scattering on two ions in a strong laser field, $\omega=0.2$ and $E_0=0.2$ at a distance of $D=47.12$ allowing for resonant absorption of energy from the laser field. The corresponding classical cut-off energies for the fast electrons are indicated by white dashed lines. Again, the laser field is polarized along the x-axis ($\varphi=0,\pi$) which is also the alignment axis of the ions.}
\label{fig:2d_2ions}
\end {figure}

As one recognizes from the single-ion scattering case, the electron density is distributed over a large space (all angles $\varphi$) which, consequently, leads to a strongly reduced intensity of electrons in a specific energy and angle interval. Especially the case of fast electrons at highest possible acceleration shows the lowest total yield of electrons in the energy spectrum. Therefore, the question occurs, if a second ion placed at a distance $D$ which allows for resonant increase of energy, leads also to a double-plateau like structure in the energy spectra with the corresponding high energies, cf. section~\ref{sec:1d_2ions}.
To answer this question, we performed TDSE simulations for such a situation. The system is displayed in the left part of figure~\ref{fig:2d_2ions}. Two ions at a distance of $D=47.12$ are radiated with a strong laser field ($\omega=0.2, E_0=0.2$). The alignment axis of both ions is parallel to the laser polarization axis and the direction of the initial momentum of the electronic wave packet, $\boldsymbol{k}_0$. The result of a TDSE simulation with $k_0=1.0$ is given in figure~\ref{fig:2d_2ions}, together with the expected classical cut-off energies for total backward (forward) scattering (white dashed lines), estimated according to the considerations for the one-dimensional model. The signal of the singly scattered electrons with optimal energy transfer from the laser field to the electron in backward direction are easily identified (classical cut-off 1). However, in forward direction ($\varphi=\pi$) the situation is different. No obvious double-plateau like distribution can be spotted and only a very weak signal of (fast) electrons is obtained for energies below the classical estimated cut-off energy (classical cut-off 2). The reason for this is the low intensity of fast electrons from the first collision. The electrons are distributed over a large spatial area and therefore only a very small fraction is colliding with the second ion. An even smaller fraction hits this ion during the ``correct'' phase of the laser field (where the electron has the maximum quiver momentum at time of collision), which is responsible for the cut-off energy.

\subsection{Focussing by external confinement}
\begin {figure}
 \begin {minipage}{0.4\textwidth}
\includegraphics[width=0.98\textwidth]{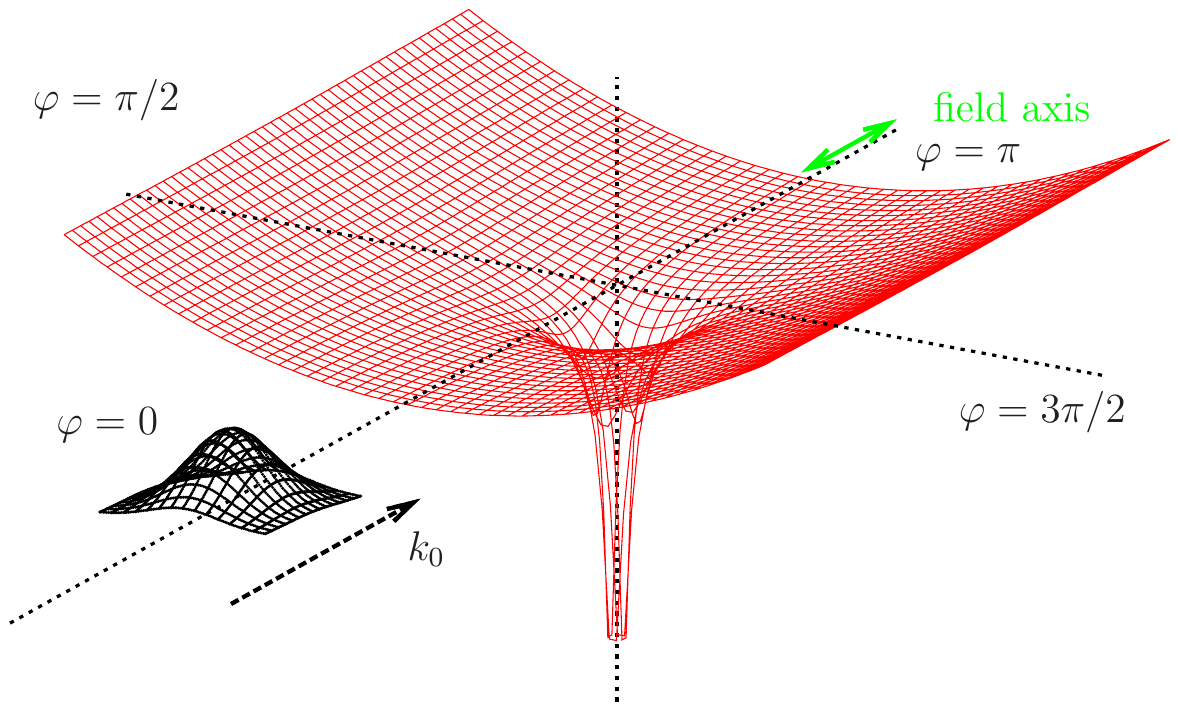}  
 \end {minipage}
 \begin{minipage}{0.59\textwidth}
 \includegraphics[width=0.98\textwidth]{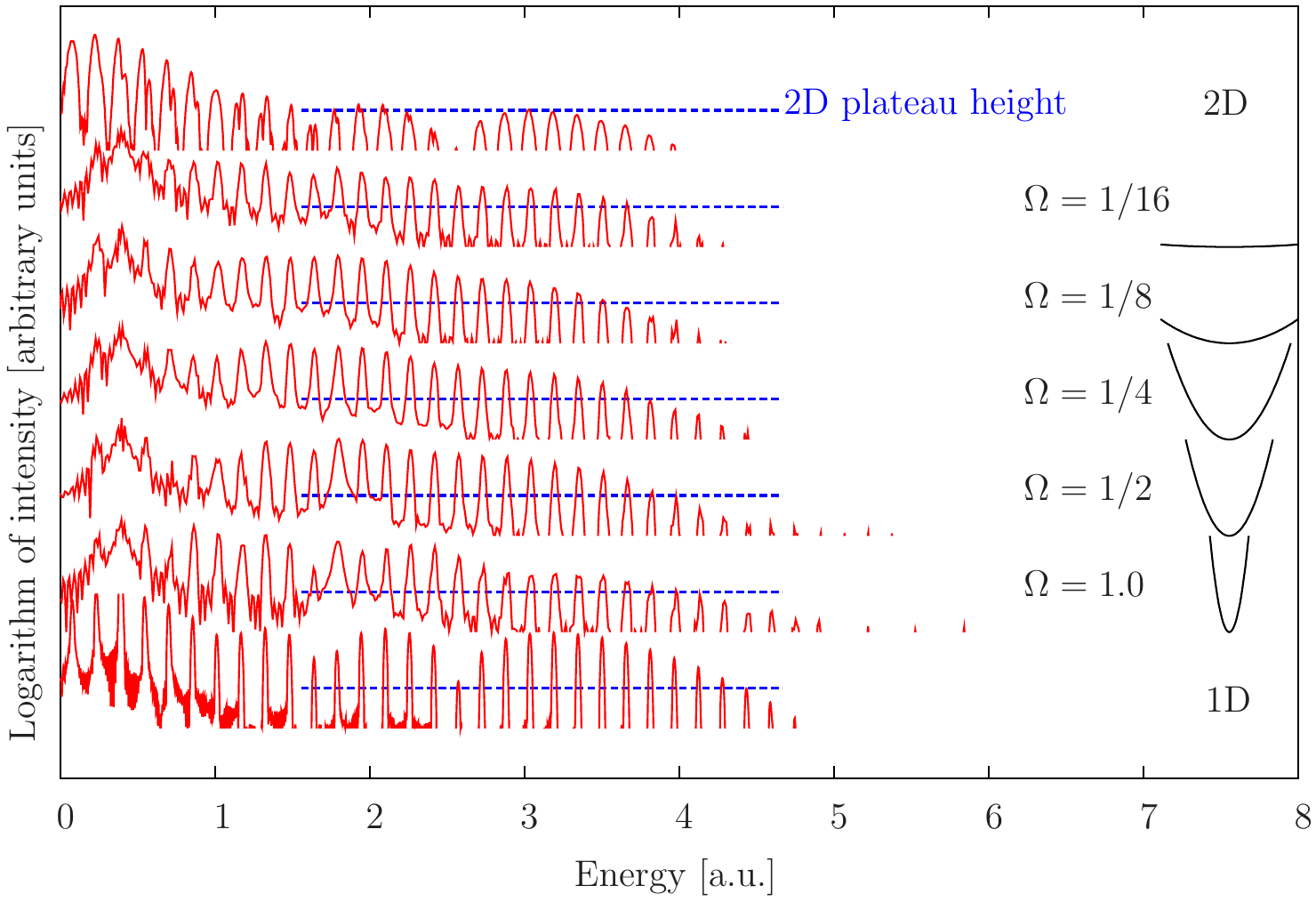}
  \end{minipage}
 \caption{(color online) 2D Coulomb scattering on a single ion in an external confinement. The confining potential, cf. equation~\eqref{eq:2d_confinement}, is perpendicular to the field axis and the propagation direction of the initial wave packet $\boldsymbol{k}_0$. The right picture shows the energy spectra of \emph{backward} scattered electrons for different confinement strengths, $\Omega$, ranging from the pure two-dimensional case ($\Omega=0$) to the one-dimensional case. The (blue) dashed line indicates the height of the intensity plateau found in the 2D calculation as a reference. }
\label{fig:2d_conf_single}
\end{figure}

To increase the intensity of optimally scattered (fast) electrons, one possibility is the introduction of additional external potentials \cite{bauchdiplom}.
The simplest idea is to confine the electron wave function in direction of the laser field, the travelling direction of the wave packet and the alignment axis of the scattering centers. This can be achieved by using a harmonic oscillator potential perpendicular to $\boldsymbol{k}_0$,
\begin{equation}
 V_{\textup{conf}} = \frac {1}{2} \Omega^2 y^2 \;.
\label{eq:2d_confinement}
\end{equation}
The trap frequency in $y$-direction is denoted by $\Omega$, the motion in $x$-direction (direction of $\boldsymbol{k}_0$ and the laser field) is not affected by this potential. The setup is schematically drawn in figure~\ref{fig:2d_conf_single}. The benefits of this potential are two-fold: first, the numerical effort decreases enormously, since in $y$-direction only a  small extension of the numerical grid is needed. The second advantage originates from the physics: the potential focusses the electron wave packet in direction of its central momentum and, therefore, onto a second scattering center, if present. This should increase the total yield in intensity of fast electrons by orders of magnitude and finally should converge (for large $\Omega$) to the results of the 1D calculations. 

A few adaptions of the methods used previously in this paper have to be made, owing to the changed geometry. First, the initial conditions are modified according to the trapping potential in $y$-direction. Along $x$, still a free travelling Gaussian wave packet with initial momentum $k_{0,x}$ is used, in $y$-direction however, the first eigentstate of the harmonic confinement, i.e. a Gaussian distribution with $\sigma_y=1/\sqrt{\Omega}$, is chosen. Secondly, the detector is adjusted to this setup. Besides an energy-shift by the eigenvalue in $y$-direction, the energy spectrum is integrated over an angle element $\Delta \varphi$ in forward ($\varphi=\pi$) and backward ($\varphi=0,2\pi$) direction since a full angular-resolved energy spectrum for such a setup is not useful. Typically $\Delta \varphi=0.1/2\pi$ is used.

The results of TDSE simulation for such a process with $k_{0,x}=1.0$ are given in Figure~\ref{fig:2d_conf_single}. Only the energy spectrum of backward scattered electrons into the angle element $-\Delta \varphi \dots + \Delta \varphi$ is plotted. All spectra show a similar structure with a cut-off energy near the expected classical maximum energy transfer. For a comparison of the intensities (logarithmic scale) the height of the backward scattering plateau for the 2D case is indicated by a blue dashed line as a guide to the eye. The trend which arises by introducing the confinement is easily observed: the intensity of backward scattered electrons is increased with stronger entrapment in $y$-direction. This is, of course, the effect of the expected focussing. The comparison with the 1D calculations (lowest graph) is difficult. The normalization of the spectrum underestimates the total yield since no corresponding angle element $\Delta \varphi$ is available in 1D. Therefore the line for the 2D plateau height indication should be located somewhat lower than given in the graph. 

In the last part we discuss the scattering on an ion pair in a harmonic confinement. The alignment axis of both ions at the resonance distance $D=47.12$, the laser polarization axis and the initial momentum $k_0$ of the wave packet are parallel. The angle-integrated (backward: $\varphi= -\Delta \varphi \dots + \Delta \varphi$, foward: $\varphi=\pi-\Delta \varphi \dots  \pi + \Delta \varphi$) energy spectrum of scattered electrons is shown in figure~\ref{fig:2d_conf_2ions}. The (strong) external confinement of $\Omega=1.0$ leads to a focussing of the backward scattered electrons during the first collision on the second ion, which now plays a more significant role in the formation of the energy spectrum. The occurrence of a double-plateau like distribution of electrons in the energy domain can be found, which is not possible in the pure 2D case without an additional external confinement ($\Omega=0$), cf. figure~\ref{fig:2d_2ions}. But, in contrast to the 1D calculations, the total yield is again very weak and only hardly distinguishable from the noise in the spectrum due to numerical effects. Additionally, the cut-off energies of the two plateaus are not as prominent as in 1D which might be attributed to the fact that electrons are scattered at different angles and, thus, different energies are collected by the confinement which smear out the distribution.

\begin {vchfigure}
  \sidecaption
 \includegraphics[width=0.5\textwidth]{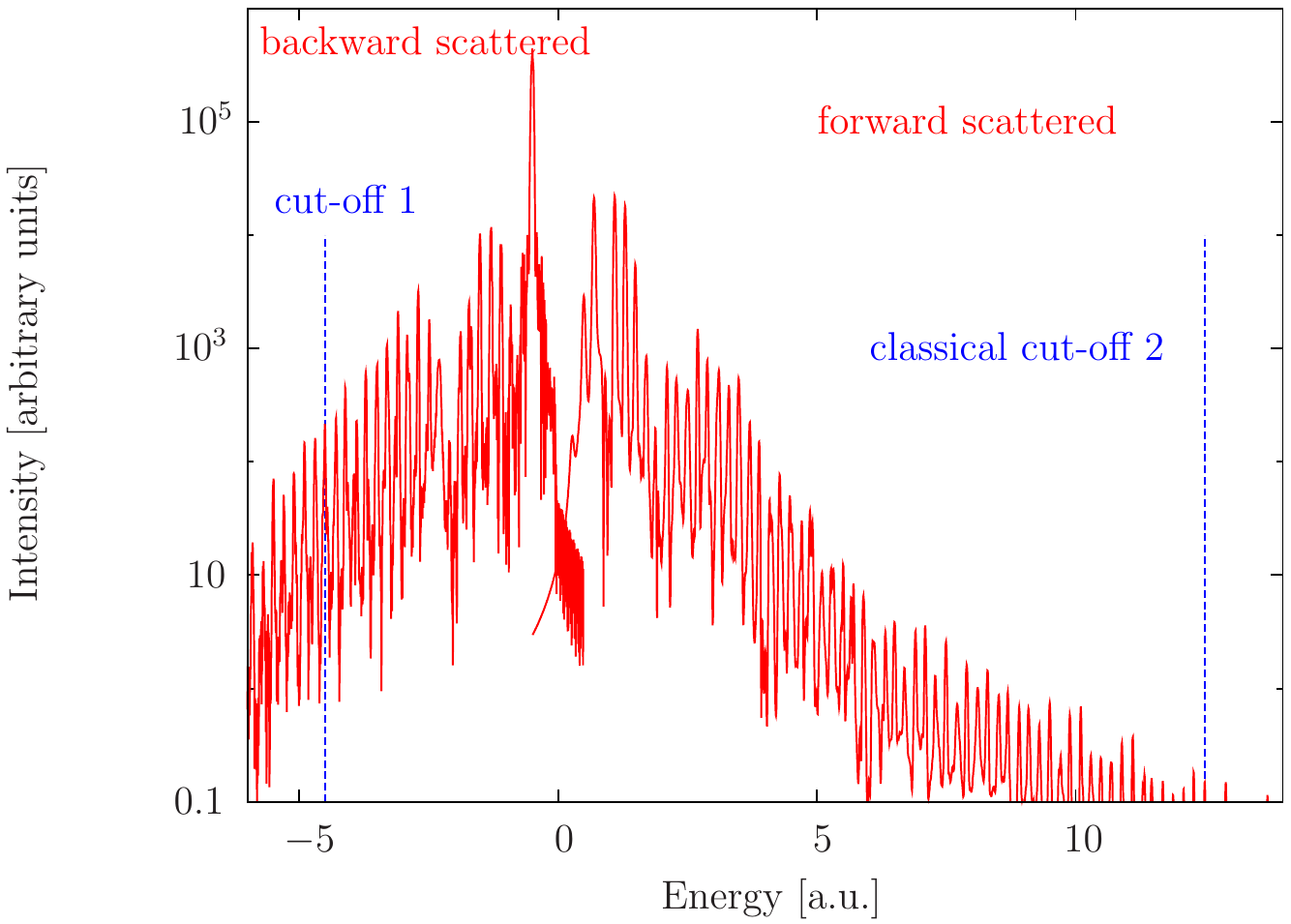}
 \caption{(color online) 2D Coulomb scattering on two spatially fixed ions at a distance $D=47.12$ in an external confinement, $\Omega=1.0$, perpendicular to the alignment axis of both ions. The setup is placed in a strong laser field with $\omega=0.2$ and $E_0=0.2$. The classically expected cut-off energies for the plateau-like distribution in the energy domain is indicated by (blue) dashed lines.}
\label{fig:2d_conf_2ions}
\end {vchfigure}

\section {Summary and Outlook}
We have investigated electron-ion collisions in strong laser fields quantum-mechanically by numerically solving the time-dependent Schr\"odinger equation. Distributions of fast electrons being accelerated by the laser field during the collisions were observed. Different spatial arrangements of (fixed-in-space) ions were considered to benefit from resonance conditions. The already known single and double ion scattering setups were generalized to three ions, where a further increase in electron energies could be demonstrated in one-dimensional model systems. For laser systems with an intensity of $I=1.4 \cdot 10^{15}\;\textup{W}/\textup{cm}^2$ at an photon energy of $5.4\textup{eV}$, the scattering on spatially correlated ions at a distance of $D \approx 2.4 \textup{nm}$ leads to a maximum energy of $340\textup{eV}$ (by using an initial electron energy of $13.6\textup{eV}$). An even higher energy can be obtained by utilizing three ions. Here one expects, according to our simulations, a maximum electron energy of $670\textup{eV}$, accelerated over a total setup which measures only approximately $4\textup{nm}$.

 To account for angular degrees of freedom in the scattering in more than one dimension and the expected decrease in efficiency, 2D systems were investigated in detail. The corresponding angular distributions of fast electrons showed a rich structure, where the spreading of the wave function and the significant decrease in intensity became visible. Therefore, resonance phenoma found in one-dimensional systems by collisions on more than one ion, could only be identified by introducing additional focussing potentials, which offer the possibility to increase the intensity of optimally scattered electrons enormously, compared to unconfined 2D systems. An additional increase of the high energy electron yield could be achieved by using highly charged ions and optimized laser polarization (pulse shaping).

\paragraph{Acknowledgements}
We acknowledge stimulating discussions with H.-J. Kull and Yu. Lozovik. This work was supported by the Innovationsfond Schleswig-Holstein.

\end{document}